\documentclass[prl,showpacs,showkeys,preprintnumbers,amsmath,amssymb,twocolumn]{revtex4-1}
\usepackage{graphicx}
\usepackage{amsfonts}
\usepackage{color}

\def\<{\langle}
\def\>{\rangle}
\def\tmb{t_w^{\rm mb}}
\begin{document}

\title{From cage--jump motion to macroscopic diffusion in supercooled liquids}

\author{Raffaele Pastore}
\email{pastore@na.infn.it}
\affiliation{CNR--SPIN, Dip.to di Scienze Fisiche,
 Universit\`a di Napoli ``Federico II'', Naples,
Italy}

\author{Antonio Coniglio}
\affiliation{CNR--SPIN, Dip.to di Scienze Fisiche,
 Universit\`a di Napoli ``Federico II'', Naples,
Italy}

\author{Massimo Pica Ciamarra}
\affiliation{CNR--SPIN, Dip.to di Scienze Fisiche,
 Universit\`a di Napoli ``Federico II'', Naples, Italy}

\date{Received: \today / Revised version: }

\begin{abstract}
The evaluation of the long term stability of a material requires the estimation
of its long--time dynamics. For amorphous materials such as structural glasses,
it has proven difficult to predict the long--time dynamics starting from static measurements.
Here we consider how long one needs to monitor the dynamics 
of a structural glass to predict its long--time features. 
We present a detailed characterization of the statistical features
of the single--particle intermittent motion of structural glasses,
and show that single--particle jumps are the irreversible events
leading to the relaxation of the system. 
This allows to evaluate the diffusion constant 
on the time--scale of the jump duration,
which is small and temperature independent, 
well before the system enters the diffusive regime.
The prediction is obtained by analyzing the particle trajectories via a parameter--free algorithm.
\end{abstract}

\pacs{64.60.ah,61.20.Lc,05.50.+q}

\maketitle

\section{Introduction}
\begin{figure}[t!]
\begin{center}
\includegraphics*[scale=0.33]{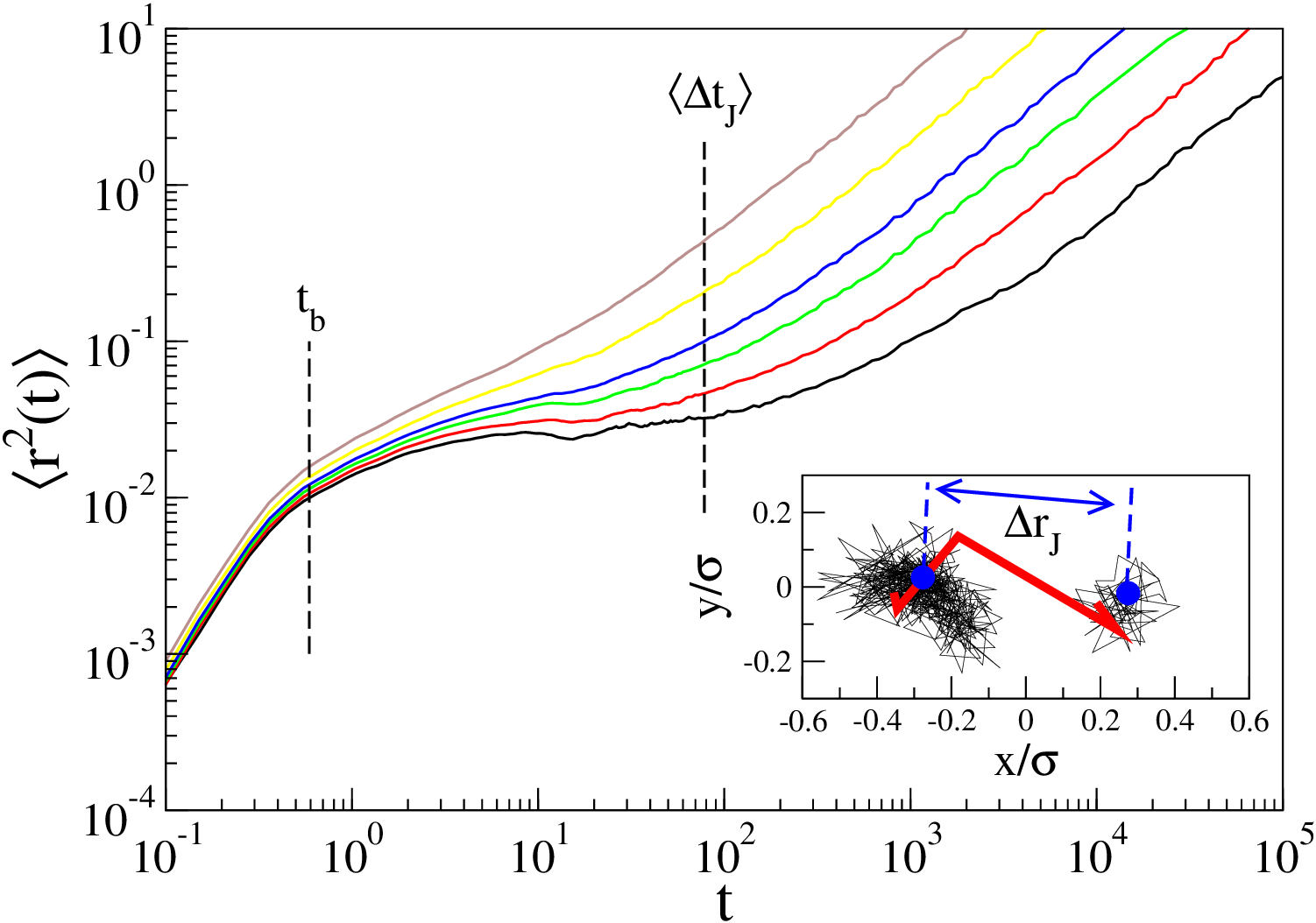}
\end{center}
\caption{\label{fig:figura1}
Mean square displacement for $T = 25,22,20,19,18$ and $17\cdot10^{-3}$, from left to right.
The dashed lines indicate the ballistic time, $t_b$, and the average time of flight of cage jumps, $\<\Delta t_J \>$.
The inset illustrates a portion of a particle trajectory at temperature $T=17\cdot10^{-3}$, that
our algorithm decomposes in two cages connected by a jump. The jump consists
of four consecutive segments, each one corresponding to the displacement of the particle in a time $\delta t = 10$ (red thick line). 
The jump length, $\Delta r_J$, is defined as the distance between the center of mass of the two cages.
}
\end{figure}

The glass transition is a liquid
to solid transition that occurs 
on cooling in molecular and colloidal systems.
The transition is characterized by a slowing down of the dynamics
which is more pronounced than that occurring in critical phenomena,
and that takes place without appreciable structural changes.
Understanding the origin of this slowdown is 
a major unsolved problem in condensed matter~\cite{Angell, Debenedetti},
that has been tackled developing different competing theories that
try to describe the observed phenomenology from a thermodynamic
or from a kinetic viewpoint. See Ref.~\cite{Biroli,StillingerDebenedetti} for recent reviews.
From a practical viewpoint, solving the glass transition problem is of interest
as this would allow to estimate the long
term stability of glassy materials, e.g. drugs and 
plastic materials such as organic solar cells~\cite{cells}.
In this respect, since we are not yet able to fully predict the long term dynamics of a glassy system from its
static properties, it becomes of interest to consider how long we need to observe a system before we can predict its dynamical features. 
This is a promising but still poorly investigated research direction.
Since the relaxation process occurs through a sequence of irreversible events, in 
this line of research it is of interest to identify these events and
to determine their statistical features. 
For instance, by identifying the irreversible events with transitions between (meta)basins of
the energy landscape~\cite{Heuer03, Heuer05, Heuer12}, that can only be detected in small enough systems ($N\lesssim 100$), 
it is possible to predict the diffusivity from a short time measurement.
Similarly, the diffusivity can also be predicted if the irreversible events are associated
to many--particles rearrangements~\cite{WidmerCooper,Procaccia,Yodh,Onuki12, Onuki13, Kawasaki}, 
that are identified via algorithms involving many parameters.
We approach this problem considering that in glassy systems particles 
spend most of their time confined within the cages formed by their neighbors, 
and seldom make a jump to a different cage~\cite{Intermittence}, as illustrated in
Fig.~\ref{fig:figura1}(inset).
This cage--jump motion is characterized by 
the waiting time before escaping a cage, 
by the typical cage size, and by the type of walk resulting from subsequent jumps.
Previous experiments and numerical studies have investigated some of 
these features~\cite{WeeksScience, WeeksPRL, Leporini_PRE, Candelier_gm, Candelier_glass, Chandler_PRX, Bingemann, Onuki12, Onuki13, Kawasaki, Chaudhuri07},
as their temperature dependence gives insight into the microscopic origin of
the glassy dynamics. 
Here we show that single--particle jumps are the irreversible events leading to the relaxation
of the system and clarify that the typical jump duration $\<\Delta t_J\>$ is
small and temperature independent: this allows to estimate 
the single particle diffusion constant resulting from a sequence of jumps, 
$D_{J}$, and the density of jumps, $\rho_{J}$, on the time scale of $\<\Delta t_J\>$,
if the size of the system is large enough.
These estimates lead to an extremely simple short time prediction of the diffusivity of the system
\begin{equation} \label{eq:main}
D(T) = D_{J}(T) \rho_{J}(T),
\end{equation}
that can be simply exploited by investigating the particle trajectories via a parameter--free algorithm.

\section{Methods}
We have obtained these results via NVT molecular dynamics simulations~\cite{LAMMPS} of a model glass former, 
a 50:50 binary mixture of $N = 10^3$ disks in two dimensions, with a diameter ratio $\sigma_{L}/\sigma_{S} =1.4$
known to inhibit crystallization, at a fixed area fraction $\phi = 1$. 
Two particles $i$ and $j$, of average diameter $\sigma_{ij}$, interact
via an Harmonic potential, $V(r_{ij}) = \epsilon \left((\sigma_{ij}-r_{ij}) /\sigma_L\right)^2$,
if in contact, $r_{ij} < \sigma_{ij}$. This interaction is suitable
to model soft colloidal particles~\cite{Likos, Berthier_EPL09, Yodh2011, Zaccarelli}.
Units are reduced so that $\sigma_{L}=m=\epsilon=k_B=1$, 
where $m$ is the mass of both particle species and $k_B$ the Boltzmann's constant.
In the following, we focus on results concerning the small particles, but analogous
ones hold for both species.
Our results rely on the introduction of a novel algorithm to identify in the particle trajectories both the cages, 
as in previous studies, as well as the jumps, whose features are here studied for the first time.
The algorithm is based on the consideration that,
for a caged particle, the fluctuation $S^2(t)$ of the position on a timescale $\delta$ corresponding
to few particle collisions is of the order of the Debye--Waller factor $\<u^2\>$.
By comparing $S^2(t)$ with $\<u^2\>$ we therefore consider a particle as caged
if $S^2(t) < \<u^2\>$, and as jumping otherwise. Practically,
we compute $S^2(t)$ as $\< (r(t)-\<r(t)\>_\delta)^2\>_\delta$,
where the averages are computed in the time interval $[t-\delta:t+\delta]$,
and $\delta=10t_b$ where $t_b$ is the ballistic time. Following Ref.s~\cite{Leporini, Leporini_JCP},
at each temperature we define $\<u^2\> = r^2(t_{DW})$,
where $t_{DW}$ is the time of minimal diffusivity of the system, i.e. the time
at which the derivative of $\log \<r^2(t)\>$ with respect to $\log(t)$ is minimal~\cite{nota_u2}.
The algorithm is slightly improved to
reduce noise at high temperatures, where cages are poorly defined due to the absence
of a clear separation of timescales~\cite{in_preparation}.
At each instant the algorithm gives access to the density of jumps, $\rho_J$,
defined as the fraction of particles which are jumping, and to the density of cages, $\rho_C=1-\rho_J$.
We stress that in this approach a jump is a process with a finite duration, as illustrated in Fig.~\ref{fig:figura1}(inset).
Indeed, by monitoring when $S^2$ equals $\<u^2\>$, we are able to 
identify the time at which each jump (or cage) starts and ends.
The algorithm is robust with respect to the choice of the time
interval over which the fluctuations are calculated,
as long as this interval is larger than the ballistic time, and much smaller
than the relaxation time.
Due to its conceptual simplicity, this algorithm is of general applicability
in experiments and simulation. Indeed, its only parameter
is the Debye-Waller factor, which is a universal feature of glassy systems.

\section{Results}
\begin{figure}[t!]
\begin{center}
\includegraphics*[scale=0.33]{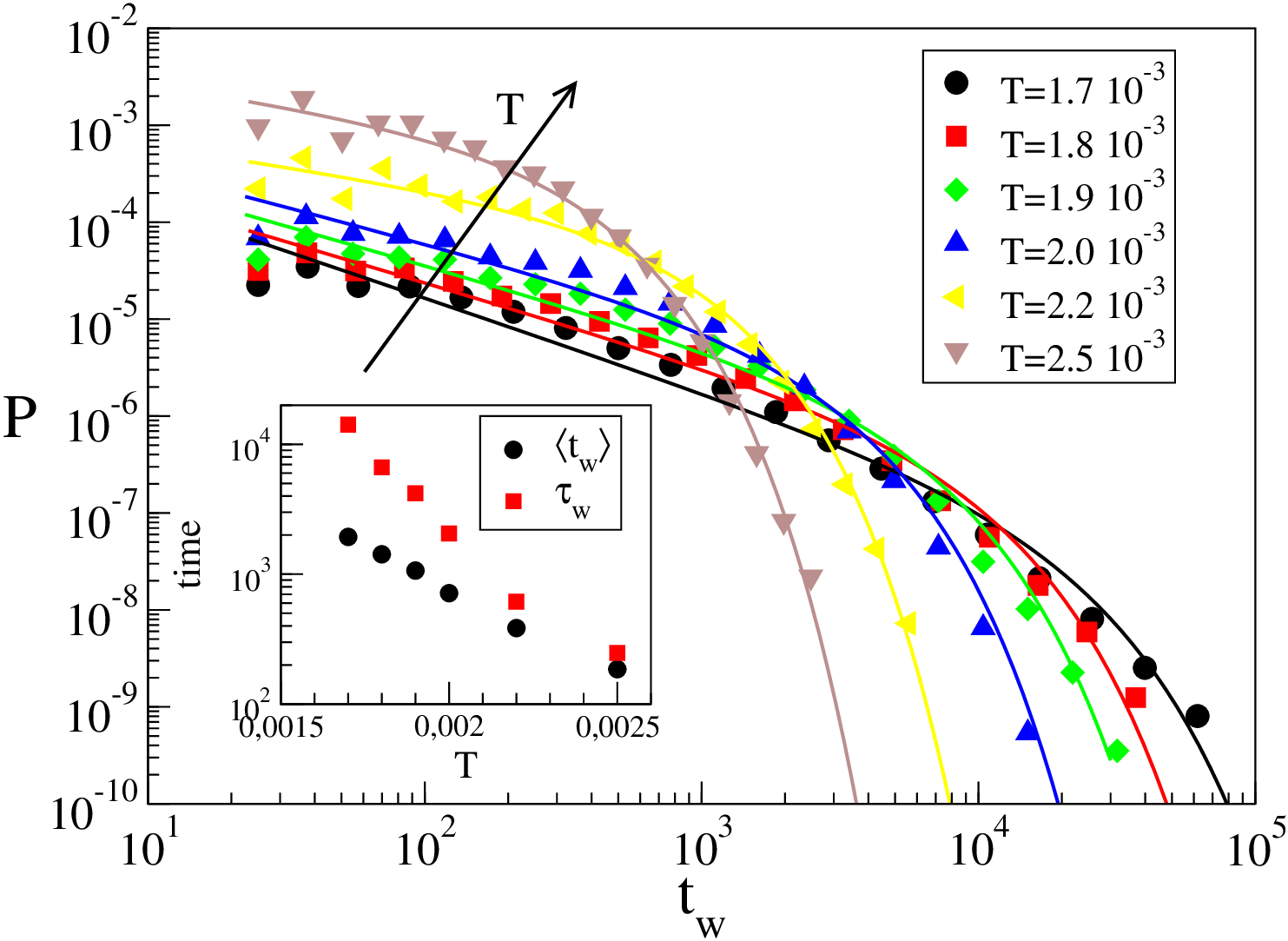}
\end{center}
\caption{\label{fig:P_tw}
Waiting time distribution, $P(t_w)$ at different temperatures,  as illustrated in Fig.~\ref{fig:figura1}.
Lines are fits to Eq.~\ref{eq:Ptw}. The inset illustrates the temperature dependence of 
the mean, $\langle t_w\rangle$, and of the cutoff time, $\tau_w$.
}
\end{figure}

\begin{figure}[t!]
\begin{center}
\includegraphics*[scale=0.33]{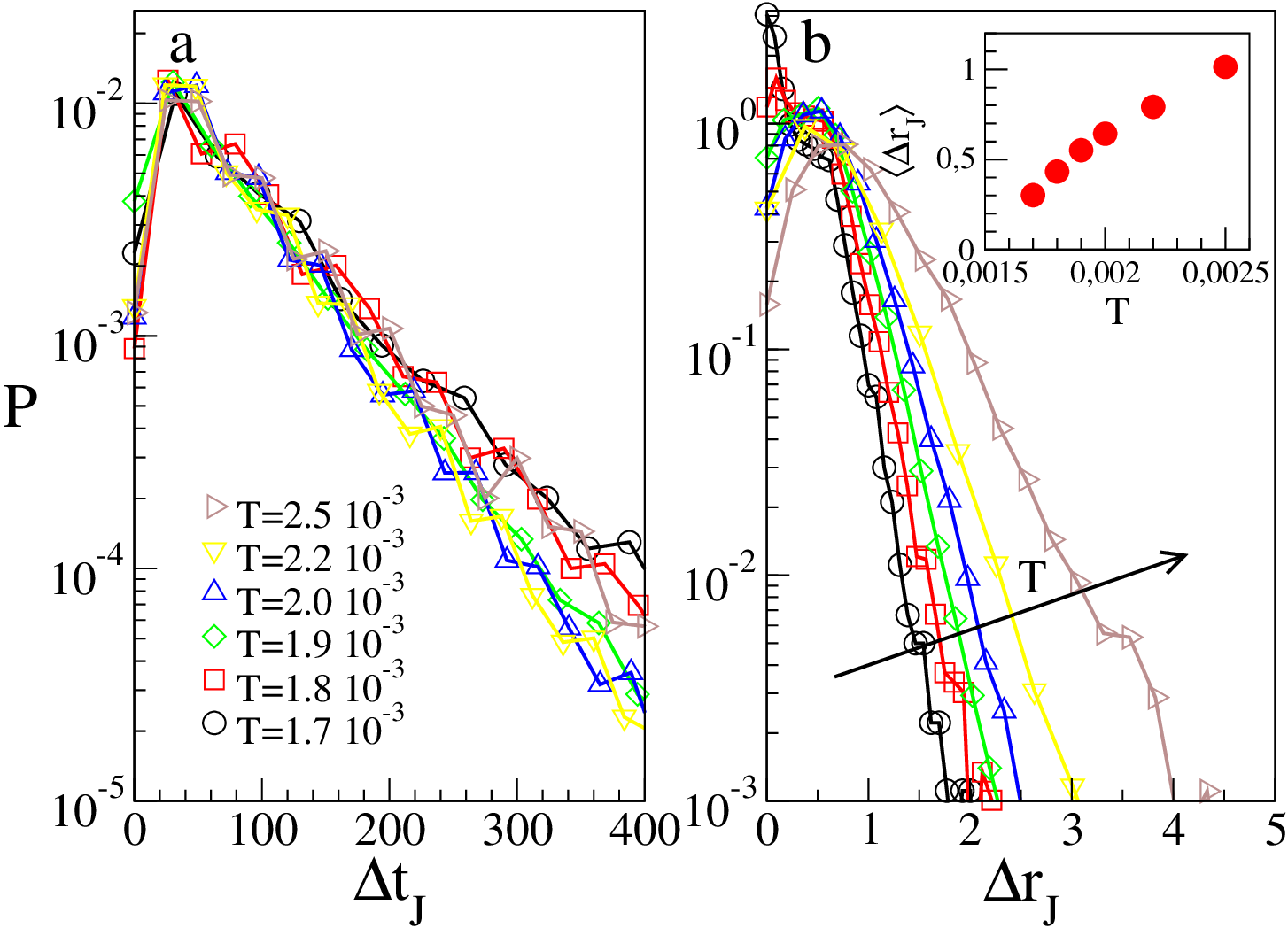}
\end{center}
\caption{\label{fig:jump_stat}
Distribution of the time of flight (a) and of the length (b) of the jumps.
Inset: temperature dependence of the averaged jump length, $\langle\Delta r_J\rangle$.
} 
\end{figure}

We have divided the trajectory of each particle in a sequence
of periods during which the particle is caged, of duration $t_w$, separated by periods 
during which the particle is jumping, of duration $\Delta t_J$.
The waiting time distribution within a cage, $P(t_w)$, illustrated in Fig.~\ref{fig:P_tw},
is well described by a 
by power law with an exponential cutoff, 
\begin{equation}
\label{eq:Ptw}
P(t_w) \propto t_w^{-\beta}\exp\left(-\frac{t_w}{\tau_w}\right),
\end{equation}
as observed in different systems~\cite{Leporini_PRE, Bingemann}. The exponent
$\beta(T)$ increases by lowering the temperature, ranging in the interval $\beta \in [0.4,0.95]$.
Since $\<t_w (T)\> = \tau_w(T) (1-\beta(T))$, this implies that
the average waiting time $\<t_w\>$ grows slower than the exponential cutoff time, $\tau_w(T)$,
as illustrated in the inset.
The time of flight distribution $P(\Delta t_J)$, illustrated in Fig~\ref{fig:jump_stat}a, 
decays exponentially. The collapse of the curves corresponding to different temperatures clarifies that, while the average time a particle
spend in a cage increases on cooling, the average duration of a jump is temperature independent. 
We find $\langle\Delta t_J\rangle \simeq 100 t_b$. 
We note that the presence of a temperature dependent waiting time and of a
temperature independent jump time is readily explained via a two well potential
analogy; indeed, the waiting time corresponds to the time of the activated
process required to reach the energy maximum, while the jump time is that
of the subsequent ballistic motion to the energy minimum.
We define the length of a jump $\Delta r_J$ as 
the distance between the center of mass of adjacent cages,
as illustrated in Fig.~\ref{fig:figura1}(inset).
Fig~\ref{fig:jump_stat}b shows that this length is exponentially distributed, with a temperature
dependent average value.

\begin{figure}[t!]
\begin{center}
\includegraphics*[scale=0.33]{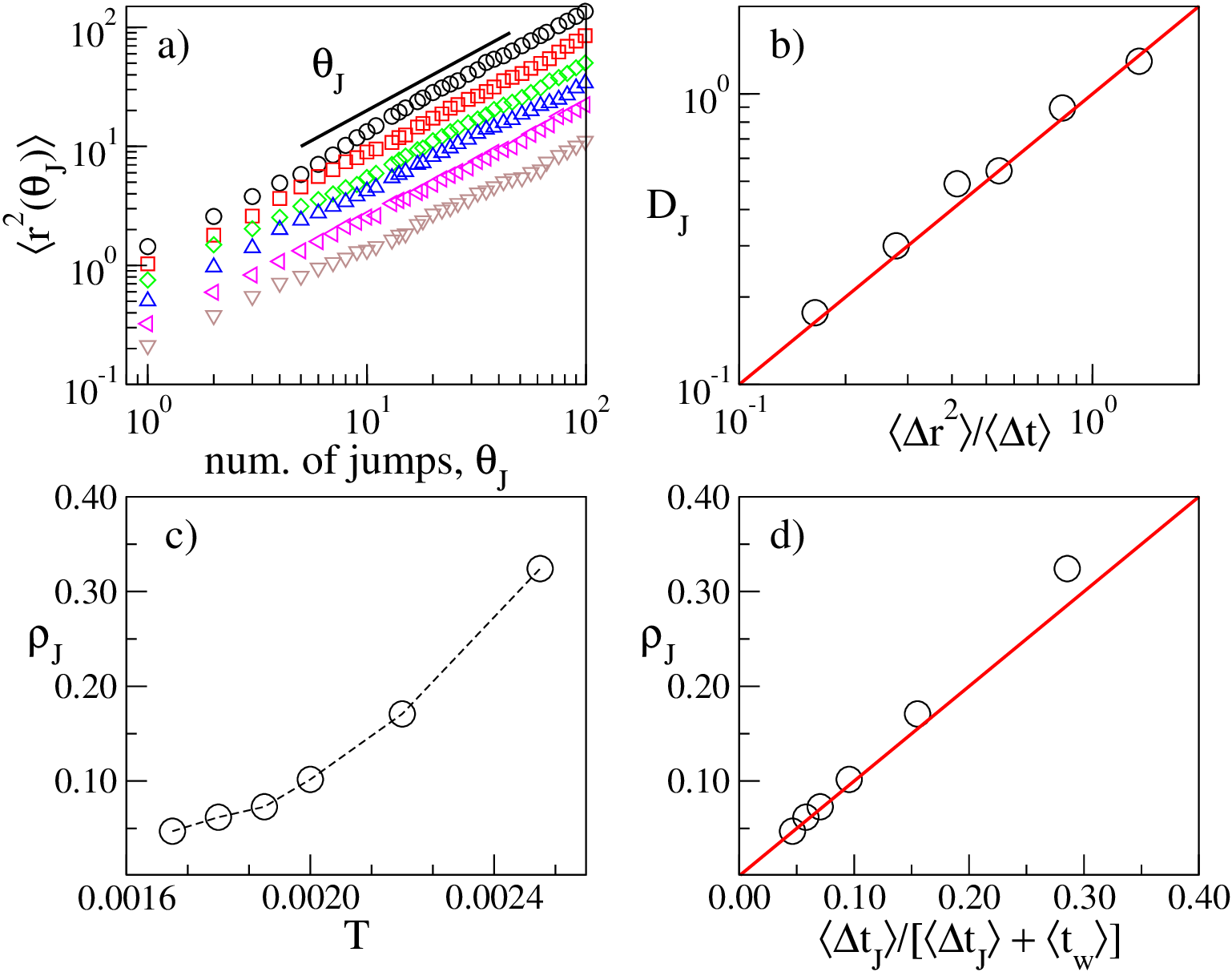}
\end{center}
\caption{\label{fig:panels}
(a) Mean square displacement as a function of the number of jumps at different temperatures, as indicated in Fig.\ref{fig:jump_stat}. 
Solid line is a power law guide to the eyes  with exponent $1$; 
(b) validation of Eq. \ref{eq:msd_nj}, that connects
the diffusion constant of the jumpers, $D_J$,  to its short time estimation,  $\frac{\<\Delta r^2 _J\>}{\<\Delta t_J\>}$. (c) Temperature dependence of the density of jumps; 
(d) validation of Eq.~\ref{eq:rho_J}, that connects the density of jumps to the timescales of the cage--jump motion.
In (b), (c) and (d), open circles and the solid line are the measured and predicted values, respectively. 
}
\end{figure}

Since the average jump length is at least a factor three larger than the cage gyration radius, which
is Gaussian distributed (not shown), one can consider each particle as a walker with a temperature dependent step size $\<\Delta r_J\>$, and 
a temperature independent time of flight $\<\Delta t_J\>$. 
The features of this walk can be inferred from the mean squared displacement $\langle r^2(\theta_J)\rangle$, 
illustrated in Fig.\ref{fig:panels}a, where the average is taken over the ensemble of particles which have performed $\theta_J$ jumps.
At all temperatures, the walk is to a good approximation diffusive from the onset.
Accordingly, we predict the diffusion constant $D_J$ of the jumpers to be that of a
pure random walk with step size $\<\Delta r_J\>$ and time of flight  $\<\Delta t_J\>$:
\begin{equation}
\label{eq:msd_nj}
D_J = \lim_{\theta_J \to \infty }\frac{\langle r^2(\theta_J)\rangle}{ \theta_J \langle\Delta t_J\rangle} = \frac{\langle\Delta r^2_J\rangle}{\langle\Delta t_J\rangle}.
\end{equation}
The validity of this prediction is verified in Fig.~\ref{fig:panels}b.
This result shows that single--particle jumps are the irreversible events leading to the relaxation of the system, 
and suggests that they are the elementary units of both 
local irreversible many--particle rearrangements~\cite{WidmerCooper,Procaccia,Yodh},
as well as of global irreversible events, such as transitions between basins in the energy landscape~\cite{Heuer03, arenzon, makse}.
In addition, Eq.~\ref{eq:msd_nj} allows to estimate a long time quantity, the jumper's diffusion constant, $D_J$, from  
properties of the cage--jump motion estimated at short times, of the order of $\< \Delta t_J\>$.
Since the time of flight $\langle\Delta t_J\rangle$ is temperature independent,
Eq.~\ref{eq:msd_nj} also clarifies that the decrease of $D_J$ on cooling is due to that of $\langle\Delta r^2_J\rangle$.
As an aside, we note that these results support the speculation of Ref.~\cite{Chaudhuri07} that rationalized data from different glass formers
in the Continuous Time Random Walk paradigm~\cite{CTRW}, postulating a simple form for the waiting
time and jump distributions. Here, we have explicitly measured the cage-jump statistical properties.

The increase of the average waiting time on cooling leads to a decrease of the density
of jumps, whose temperature dependence is illustrated in Fig.~\ref{fig:panels}c. 
Indeed, these two quantities are related 
as $\rho_J$ is to good approximation equal to the fraction of the total time particles spend jumping,
\begin{equation}
\label{eq:rho_J} 
\rho_J = \frac{\langle\Delta t_J\rangle}{\langle t_w\rangle+\langle\Delta t_J\rangle},
\end{equation}
as illustrated in Fig.~\ref{fig:panels}d. We note that 
the r.h.s. of the above equation is computed after having determined the waiting time distribution, 
i.e. on a temperature dependent timescale of the order of the relaxation time,
whereas the l.h.s.  is estimated on the small and temperature independent timescale, $\< \Delta t_J\>$.
We note, however, that $\rho_J$ can be estimated on a time scale of $\<\Delta t_J\>$
only if jumps are observed on that time scale, i.e. only if $\rho_J N>1$.
This is always the case in the investigated temperature range, as we find $\rho_J N\simeq25$ at the lowest temperature.
In general, the time of observation required to measure $\rho_J$
scales as $\Delta t=\< \Delta t_J\>/\rho_J N$. This is always
much smaller than the relaxation time, as
Eq.~\ref{eq:rho_J} leads to $\Delta t \simeq (\<t_w\>+\langle\Delta t_J\rangle)/N \ll \<t_w\> \ll \tau_w$.

\begin{figure}[t!]
\begin{center}
\includegraphics*[scale=0.33]{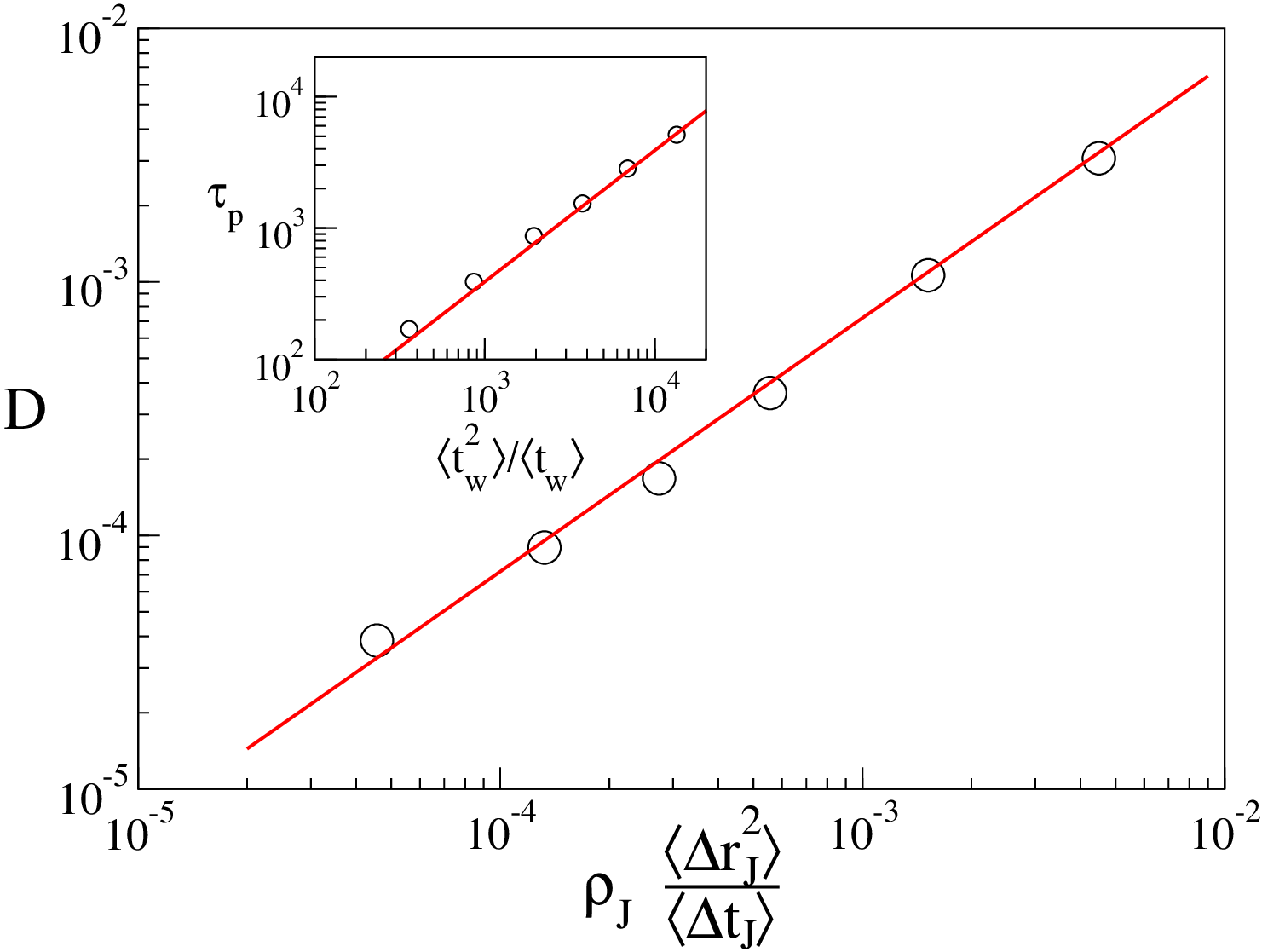}
\end{center}
\caption{\label{fig:micro_macro}
Linear dependence of the diffusion constant on features of the cage--jump motion.  
Open circles and the solid line are measured data and the prediction from Eq.~\ref{eq:diffusion}, respectively.
We stress that $D$ is estimated at long times, while $\rho_J \frac{\langle\Delta r^2_J\rangle}{\langle\Delta t_J\rangle}$
is estimated at short times, well before the system enters the diffusive regime. The solid line has slope $m \simeq 0.75$. 
Inset: the persistence relaxation time, $\tau_p$, is proportional to the ratio, $\<t_w^2\>/\<t_w\>$, of the moments of the waiting time distribution.
The solid line is a power law with exponent $1$.
} 
\end{figure}

The features of the cage--jump motion allow to predict the macroscopic diffusion via Eq.~\ref{eq:main}, 
$D = \rho_J D_J$. This equation is recovered as
\begin{equation}
\label{eq:D1}
D= \lim_{t\rightarrow \infty} \frac{1}{Nt}  \sum_{p=1} ^{N} [r_p(t) - r_p(0)]^2 = \frac{1}{Nt}  \sum_{p=1} ^{N}  \theta_J^{(p)}(t) D_J \langle\Delta t_J\rangle, 
\end{equation}
where the last equality is obtained considering that, at time $t$,
the contribution of particle $p$ to the overall square displacement is due to
$\theta_J^{(p)}(t)$ jumps of average size $D_J \langle\Delta t_J\rangle$.
Eq.~\ref{eq:main} follows as 
$\frac{1}{N} \sum_{p=1} ^{N}  \theta_J^{(p)}(t)$ is the average number
of jumps per particle at time $t$, $\<\theta_J(t)\> = \frac{t}{\langle\Delta t_J\rangle + \<t_w \>}$,
a quantity related to $\rho_J$ by Eq.~\ref{eq:rho_J}.
Eq.~\ref{eq:main} can also be expressed as 
\begin{eqnarray}
\label{eq:diffusion}
D=\rho_J \frac{\langle\Delta r^2_J\rangle}{\langle\Delta t_J\rangle}
\end{eqnarray}
thorough Eq.~\ref{eq:msd_nj}.
Our numerical results are consistent with this prediction,
as we find $D = m \rho_J \frac{\langle\Delta r^2_J\rangle}{\langle\Delta t_J\rangle}$, 
with $m \simeq 0.75$, as illustrated in Fig.~\ref{fig:micro_macro}.
We explain the value $m < 1$ considering that the time of flight, $\Delta t_J$, is a slightly underestimation
of the time required to move by $\Delta r_J$, as after jumping a particle rattles in the cage before reaching
its center of mass. 
Eq.~\ref{eq:diffusion} has two important merits. First, it connects a macroscopic property, the diffusion coefficient,
to properties of the cage--jump motion. Second, it connects a quantity evaluated in the
long time limit, to quantities evaluated at short times. 
This demonstrates that the diffusion constant can be predicted
well before the system enters the diffusive regime.
Eq.~\ref{eq:diffusion} also clarifies that two mechanisms contribute to the slowing down of the dynamics. 
On the one side $\rho_J$ decreases, as the mean cage time increases. On the other
side the diffusion coefficient $D_J$ decreases, as the jump size decreases on cooling.

We note that a previous short time prediction of the diffusion constant~\cite{Onuki13} 
was obtained identifying irreversible events with 
complex structural changes involving many-particles,
whereas our approach relies on a simple single particle analysis.
Other approaches are also not able to give a short time prediction of the diffusivity. For example, 
in order to compute the Green-Kubo integral of the velocity autocorrelation function (VACF),
one need to wait VACF to vanish, i.e. a process occurring on  a time-scale much longer than the jump duration.
In addition, the VACF approach requires to estimate the particles
velocities, that is a very problematic task from the experimental viewpoint.

We now consider how jumps are related to the relaxation of the system,
that we have monitored through the persistence correlation function:
at time $t$, this is the fraction of particles that have not yet performed a jump~\cite{Chandler_PRE, Berthier_book, PRL2011}.
From the decay of this correlation function we have estimated the persistence relaxation
time, $\tau_p$ ($p(\tau_p)=e^{-1}$), we have found to scale as the decay time of the waiting time distribution, $\tau_p \propto \tau_w$,
not as the average waiting time $\<t_w\>$.
This is explained considering the spatial heterogeneity of the dynamics.
Indeed, in the system there are mobile regions
that last a time of the order of the relaxation time~\cite{in_preparation,WeeksScience, Chaudhuri07, Chaudhuri}
where the typical waiting time is smaller than the average.
The subsequent jumps of particles of these regions influence the average waiting
time $\<t_w\>$ but do not contribute to the decay of the persistence correlation function,
which is therefore controlled by the decay time of the waiting time distribution, $\tau_p \propto \tau_w$.
It is also possible to relate
$\tau_p$ to the first two moments of $P(t_w)$, as due to Eq.~\ref{eq:Ptw}
$\tau_w \propto \<t_w^2\>/\<t_w\>(2-\beta) \simeq \<t_w^2\>/\<t_w\>$ (see Fig.~\ref{fig:micro_macro}, inset).
This expression for the relaxation time, and Eq.~\ref{eq:diffusion} for the diffusion 
coefficient, are formally analogous to those suggested by trap models~\cite{Bouchaud},
that interpret the relaxation as originating from a sequence of jumps
between metabasins of the energy landscape~\cite{Heuer03, arenzon, makse}.
Indeed, trap models predict the diffusion coefficient and the persistence relaxation time~\cite{Chandler_PRE, Berthier_book, PRL2011}
to vary as $D \propto a^2/\langle \tmb \rangle$, and as 
$\tau_p \propto \langle (\tmb)^2 \rangle / \langle \tmb \rangle$. Here $\tmb$ is the waiting
time within a metabasin, and $a$ the typical distance between two adjacent metabasins in configuration space.
It is therefore worth stressing that, since our results concern the single particle intermittent motion,
they have a different interpretation and a different range of applicability.
In particular, since $\<\tmb\>$ varies with system size as $O(1/N)$, transitions between metabasins can only be revealed 
investigating the inherent landscape dynamics of small ($\sim$100 particles)
systems~\cite{Heuer03}, and models to infer the dynamics in the thermodynamic limit need to be developed~\cite{Heuer05, Heuer12}.
Conversely, our prediction for the diffusion coefficient lacks any system size dependence
and works at short times, as previously discussed.
These results support a physical interpretation of the relaxation in terms of trap models,
but clarify that it is convenient to focus on single particle traps, 
rather than on traps in phase space, at least as long as the relaxation process occurs via
short-lasting jumps.

\section{Discussion}
We have shown that the diffusion coefficient of a glass former
can be estimated on a small timescale, which is of the order
of the jump duration and much smaller that the time
at which the system enter the diffusive regimes if the system size is large enough, $\rho_J N > 1$.
This is so because jumps are irreversible events.
This prediction requires the identification of cages and jumps
in the particle trajectories, we have show to be easily determined
via a parameter-free algorithm if 
cages and jumps are characterized by well separated time scales.
This result is expected to be relevant in real world applications
in which one is interested in 
predicting the diffusivity of systems that are in equilibrium or in a stationary state.
It can also be relevant to quickly determine an upper bound
for the diffusivity of supercooled out--of--equilibrium systems.

Open questions ahead concern the emergence of correlations between jumps of a same particle
closer to the transition of structural arrest, and the presence of spatio--temporal correlations 
between jumps of different particles.
In addition, we note that persistence correlation function behaves analogously to a self--scattering correlation function at a wavevector 
of the order of the inverse jump length. In this respect, a further research
include the developing of relations between the features of the cage--jump motion,
and the relaxation time at different wave vectors.

\begin{acknowledgments}
We thank MIUR-FIRB RBFR081IUK for financial support.
\end{acknowledgments}


\begin{thebibliography}{30}
\bibitem {nota_u2}
The error in the estimation of $\<u^2\>$ is small, of the order of 3\%, as in
the subdiffusive regime the mean square displacement does not vary much with time.
\bibitem{notadr}{Since jumps do not start/end from the center of mass of a cage,
 the length of a jump length $\lambda$ differs from $\Delta r_J$ by a quantity of the order of the cage gyration radius. 
 We consider $\lambda = \Delta r_J$ as errors made in the estimation of the length of successive jumps cancel out
 when computing the overall particle displacement.}
\end{thebibliography}
\end{document}